\documentclass[sigconf,nonacm]{acmart}

\usepackage{booktabs}
\usepackage{balance}
\usepackage{enumitem}   
\usepackage{tikz}
\usetikzlibrary{positioning,arrows.meta}

\begin{document}

\title{SCOUT: Supply-Aware Cold-Start Proactive Query Suggestion for Travel Search}

\author{Hao Li}
\email{hao.li@airbnb.com}
\affiliation{%
  \institution{Airbnb, Inc.}
  \country{USA}
}

\author{Shashank Reddy}
\email{shashank.reddy@airbnb.com}
\affiliation{%
  \institution{Airbnb, Inc.}
  \country{USA}
}

\author{Kedar Bellare}
\email{kedar.bellare@airbnb.com}
\affiliation{%
  \institution{Airbnb, Inc.}
  \country{USA}
}

\author{Ashish Jain}
\email{ashish.jain@airbnb.com}
\affiliation{%
  \institution{Airbnb, Inc.}
  \country{USA}
}

\author{Stephanie Moyerman}
\email{stephanie.moyerman@airbnb.com}
\affiliation{%
  \institution{Airbnb, Inc.}
  \country{USA}
}

\renewcommand{\shortauthors}{Li et al.}

\begin{abstract}
Generative query suggestion, powered by Large Language Models (LLMs), has become
increasingly popular in search and conversational systems to reduce user friction and
guide intent formulation. Existing approaches align suggestions with user preferences
(e.g., clicks or conversions). This works for open-ended applications like chatbots and
personal assistants, where the result space is unconstrained or historical user
free-text queries are abundant.

However, applying these methods to travel search presents two limitations. First, travel
search is fundamentally constrained by physical inventory; a query (e.g., ``romantic
beachfront villa'') may yield abundant results in Bali but few in Tokyo, so aligning with
user preferences is not by itself grounded in what can be offered. Second, travel
platforms traditionally rely on faceted search interfaces with no free-text queries. This
creates a cold-start problem: without historical query logs there is no demand-side
data for alignment, and without a seed query at request time, suggestions must be
generated proactively from structured context alone.

To address these challenges, we propose SCOUT, a bootstrapping framework for
supply-aware proactive query suggestion. SCOUT overcomes the data gap by substituting
missing demand-side user feedback with supply-side system feedback. It treats the
search engine as a reinforcement learning environment, deriving a dense reward from the
production reranker's query-listing match scores, and optimizes the policy with Group
Relative Policy Optimization (GRPO). SCOUT improves inventory match rate (IMR@18) by
12.3\% while preserving diversity, matching a compute-intensive best-of-8 policy at
zero marginal inference cost and making supply-aware suggestion deployable on a
real-time travel search path.
\end{abstract}



\acmConference[GRAIL @ CIKM '26]%
  {GRAIL 2026: Workshop on Generative, Retrieval-augmented, and Agentic
  Intelligence for Personalization}%
  {November 8, 2026}{Rome, Italy}
\acmYear{2026}
\acmDOI{}

\settopmatter{printacmref=true}
\maketitle

\makeatletter
\fancypagestyle{standardpagestyle}{%
  \fancyhf{}%
  \renewcommand{\headrulewidth}{\z@}%
  \renewcommand{\footrulewidth}{\z@}%
  \fancyhead[LE,RO]{\@headfootfont\acmConference@shortname,
    \acmConference@date, \acmConference@venue}%
  \fancyhead[LO]{\@headfootfont\shorttitle}%
  \fancyhead[RE]{\@headfootfont\@shortauthors}%
}
\pagestyle{standardpagestyle}
\makeatother


\section{Introduction}
\label{sec:intro}

Search systems are undergoing a fundamental paradigm shift, evolving from rigid keyword
matching and faceted filtering toward fluid, natural language interfaces. In this context,
Generative Query Suggestion, powered by Large Language Models (LLMs), has emerged as a
powerful tool to reduce user friction and proactively guide search intent formulation.
State-of-the-art approaches for aligning these generative models heavily rely on
Supervised Fine-Tuning (SFT) and Reinforcement Learning from Human Feedback
(RLHF)~\cite{min2025ctrguided,bytedance2025clicks,chen2026localsug}. However, these
standard methodologies implicitly assume the availability of massive demand-side datasets,
specifically historical free-text queries and user click-through logs.

Applying these standard alignment techniques to travel search introduces two challenges. First is a critical
cold-start data gap.
Historically, travel platforms have relied on highly structured UI
elements, requiring users to input destinations, travel dates, and guest counts via
calendars and dropdowns. Consequently, there is an absence of historical
free-text query logs to serve as training data. Second, travel search operates under
strict physical inventory constraints. Unlike open-domain search where any
linguistically coherent response is valid, a travel query suggestion sets an expectation of
physical inventory. Generating an appealing but unfulfillable query (e.g., suggesting a
``romantic beachfront villa'' in a landlocked city) leads to empty search result pages,
degrading user experience. Models
without supply-awareness actively hallucinate demand.

Naive approaches to supply-aware generation fail to meet production requirements.
Simple template-based generation reads as rigid and robotic, lacking the ability to
synthesize fluidly with specific travel contexts. Inventory-conditioned prompting
is also a non-starter, as a compact, latency-friendly summary of available inventory for
arbitrary travel search does not exist at generation time. A serving-time
best-of-$N$ strategy---generating multiple candidate suggestions and querying the live
search engine to serve the ones with highest inventory---solves the inventory
problem but creates a system bottleneck. Issuing $N$ heavy search requests per
interaction introduces latency and computational costs that are prohibitive for a
real-time travel search engine.

To bridge the gap between generative intent and physical inventory---without historical user data or serve-time latency and cost penalties---our core insight is to substitute missing demand-side user feedback with supply-side system feedback. We operationalize this insight through the following main contributions:
\begin{itemize}[leftmargin=*, topsep=2pt, itemsep=2pt]
\item \textbf{Problem Formulation:} We identify the unique challenge of bootstrapping free-text query suggestions for travel search. Due to the absence of historical logs and seed queries, coupled with physical inventory constraints, we formulate this task as proactive query generation directly from search contexts. (\S\ref{sec:related})
\item \textbf{Framework:} We propose SCOUT, a bootstrapping framework that leverages the live search engine as an RL training environment. It enables UX-driven evaluation without demand-side data and online RL training driven by supply-side feedback. (\S\ref{sec:bootstrap})
\item \textbf{Results:} We demonstrate that aligning models via supply-side RL achieves superior supply-awareness while maintaining linguistic diversity compared to existing baselines. Crucially, SCOUT amortizes the prohibitive costs of serve-time validation directly into the model's weights, enabling production deployment. (\S\ref{sec:results})
\end{itemize}

\section{Problem Setting \& Related Work}
\label{sec:related}

\begin{table}[t]
\centering
\small
\caption{Travel search is uniquely inventory-constrained and lacks user-authored free-text queries. Missing historical logs rule out demand-side alignment, and missing request-time inputs rule out seed-based query reformulation.}
\label{tab:setting}
\begin{tabular}{lccc}
\toprule
 & inventory- & \multicolumn{2}{c}{user-authored free-text query} \\
 \cmidrule(l){3-4}
 & constrained & as input & in logs \\
\midrule
web search / chatbot & no  & yes & yes \\
product search       & yes & yes & yes \\
travel search        & yes & no  & no  \\
\bottomrule
\end{tabular}
\end{table}

\textbf{The Travel Search Setting.}
Travel platforms have historically relied on structured, faceted search interfaces (date
pickers, destination autocomplete) rather than an open text box.
Introducing free-text search therefore starts from a setting with two properties
that, together, are unique among the domains where generative query suggestion has
been studied (Table~\ref{tab:setting}). First, there is no user-authored free-text
query: not as input at request time, and not in historical logs. Second,
travel search is bounded by physical, location-specific inventory. These two
properties are what rule out the two established families of prior work.

\textbf{Gap 1: Alignment Requires Demand-Side Data.}
Large Language Models have shifted the query suggestion space toward generative
query suggestion, where supervised fine-tuning and RLHF align suggestions with user
intent using historical free-text queries and engagement feedback
\cite{min2025ctrguided,bytedance2025clicks,chen2026localsug,guo2026onesug}. These methods assume
the demand-side data that our setting does not have.

\textbf{Gap 2: Retrieval Feedback Requires a Seed Query.}
A second family uses the retrieval system as the reward source, optimizing
query reformulation, rewriting, or expansion against retrieval outcomes
\cite{nogueira2017reformulation,agrawal2023rlqr,peng2024beque,yang2025aqe,jiang2025deepretrieval,zhang2026reflex}.
While we share the intuition of using the search engine as an RL environment, these
existing methods are fundamentally \emph{reactive transformations}. They rely on a
user-authored seed query as a reference anchor, optimizing how well the rewrite
preserves and expands that existing intent. In contrast, our task is \emph{proactive
generation}. Because the free-text query is strictly the output rather than the input,
we have no reference intent to anchor the reward. Instead of preserving intent, our
model must proactively construct plausible intents directly from structured context
and ground them in retrievable inventory.

\textbf{Formulating Proactive Generation.}
We define the task as follows: a search context $c = (\ell, d, g)$ is the
structured input travel search already accepts---destination $\ell$, date range $d$,
guest count $g$. Given $c$, a policy $\pi_\theta$ generates a slate of free-text
suggestions $S = \{s_1,\dots,s_{|S|}\}$, surfaced \emph{proactively}, before the user
issues any free-text input, since free-text search is a capability the interface has
never offered and users must be shown it to adopt it.

\section{Methodology}
\label{sec:bootstrap}

To tackle the absence of demand-side data and the physical constraints of travel
search, we introduce a supply-aware framework for generative query suggestion.
First, we establish a scalable, label-free evaluation suite to objectively measure
inventory and diversity (\S\ref{sec:bootstrap:eval}). Second, we utilize
the live search engine for supply-side feedback, training the policy via online
RL (\S\ref{sec:method:rl}).
\subsection{Evaluation without Demand-Side Data}
\label{sec:bootstrap:eval}

\textbf{Two axes: inventory and diversity.} Click-through and conversion are unavailable
during bootstrapping. To overcome this, we walk backward from the ideal guest experience
to define proxy criteria for success. A suggestion sets an expectation about the page behind it,
which gives the first axis, \emph{inventory}: clicking must open results that match the
suggestion, not an empty or low inventory page. The second axis is \emph{diversity},
both a UX necessity to prevent repetitive slates and a crucial guardrail against
diversity collapse, a documented failure mode of preference
optimization~\cite{razin2025displacement,ren2025squeezing}.

\begin{figure}[t]
\centering
\begin{tikzpicture}[
  font=\scriptsize,
  bx/.style={draw, rounded corners=1.5pt, align=center, inner sep=2.5pt,
             minimum width=54pt, minimum height=18pt},
  el/.style={font=\tiny, inner sep=1.5pt, align=center},
  ar/.style={-{Latex[length=3.5pt]}, semithick}
]
\node[bx] (ctx) {search context $c$\\from logs};
\node[bx, right=32pt of ctx] (sug) {slate $S$ of\\suggestions $s_i$};
\node[bx, right=46pt of sug] (inv) {top-$K$ listings\\$l_{i,j}$ per $s_i$};
\draw[ar] (ctx) -- node[el, above] {policy $\pi_\theta$} (sug);
\draw[ar] (sug) -- node[el, above] {retrieve from\\search engine} (inv);

\node[bx, below=15pt of sug] (div) {embed \& cluster\\$\to\ \text{distinct}\%@0.9$};
\node[bx, below=15pt of inv] (judge) {LLM judge $\hat m$\\$\to\ \text{IMR@18}$};
\draw[ar] (sug) -- node[el, right] {diversity} (div);
\draw[ar] (inv) -- node[el, right] {inventory} (judge);
\end{tikzpicture}
\Description{Flow diagram. A search context from historical logs feeds the policy, which produces a slate of suggestions. Each suggestion is sent to the search engine, which returns its top-K listings. Two scoring branches follow: the suggestion text is embedded and clustered to compute distinct percent at 0.9 for diversity, and the retrieved listings are scored by an LLM judge to compute IMR@18 for inventory.}
\caption{Label-free evaluation pipeline. Historical logs provide the context to
generate suggestion slates. We evaluate diversity via embedding-based clustering,
and inventory by retrieving the top 18 listings and scoring their
semantic match using a calibrated LLM judge.}
\label{fig:evalflow}
\end{figure}
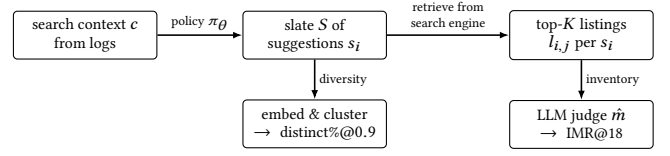

\textbf{Two metrics, one per axis.} Because no free-text queries exist, we use
historical parameterized searches as contexts and score the slate each policy generates
from them (Figure~\ref{fig:evalflow}). For inventory we define the \emph{Inventory Match
Rate}, IMR@18. For every suggestion $s_i$ in a slate $S$ we retrieve the top $K = 18$
listings $l_{i,1},\dots,l_{i,K}$, the first page of results, and average a judge's
per-listing attribute match ratio $\hat m(s_i, l_{i,j}) \in [0,1]$ over the slate:
{\scriptsize
\[
\text{IMR@18}(S) \;=\; \frac{1}{|S|\,K} \sum_{i=1}^{|S|} \sum_{j=1}^{K}
\hat m(s_i, l_{i,j}).
\]
}%
For diversity we track $\text{distinct}\%@0.9$, corpus-level distinctness per
seed: we embed each suggestion with MiniLM~\cite{wang2020minilm}, greedily
leader-cluster at cosine similarity $0.9$, and report $100 \times |\text{clusters}| /
|\text{suggestions}|$.

\textbf{A calibrated judge makes inventory measurable at scale.} Because human annotation
is unscalable, we rely on LLM judges \cite{zheng2023judge}. We first hand-curate a
representative ``golden set'' of query-to-listing matches labeled by human experts. We
then construct an LLM judge and iteratively refine its prompt against this golden set.
Once the judge's agreement with human labels reaches a pre-set threshold, we deploy
it to compute IMR@18 automatically. This calibrated pipeline provides a rigorous,
scalable, and label-free compass for evaluating our models prior to online deployment.

\subsection{SCOUT: Supply-Aware Optimization via Online RL}
\label{sec:method:rl}

\begin{figure}[t]
\centering
\includegraphics[width=0.8\linewidth]{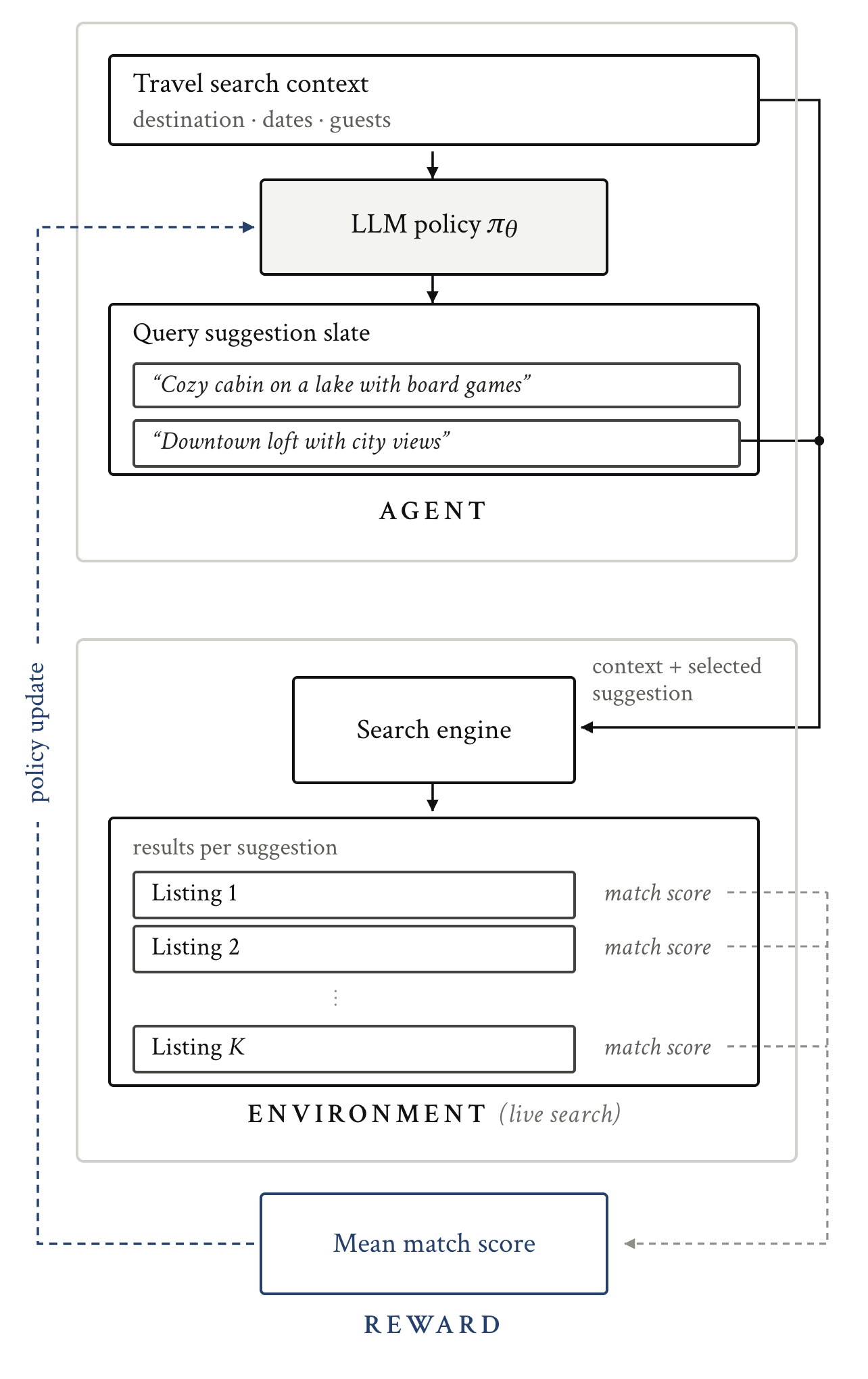}
\Description{System diagram with three parts. Agent: a travel search context of destination, dates, and guests is given to the LLM policy, which outputs a slate of query suggestions, for example "Cozy cabin on a lake with board games" and "Downtown loft with city views". Environment: the context and a selected suggestion are sent to the live search engine, which returns listings 1 to K, each with a match score. Reward: the mean match score is fed back to the policy as a policy update.}
\caption{Overview of SCOUT. The LLM generates query suggestions, which are executed against the
live search engine; the returned inventory yields the reward, and GRPO internalizes
supply-awareness into the policy weights.}
\label{fig:overview}
\end{figure}

\textbf{The Search Engine as a Reward Environment.}
The best-of-$N$ policy of \S\ref{sec:intro} can guarantee supply-backed suggestions, but it
pays for that guarantee at every impression: $N$ search calls per slate, none of which leave
the underlying model any more supply-aware. The cost recurs indefinitely because the
knowledge stays outside the model. SCOUT instead pays it once, at training time, by treating
the live search stack as a reinforcement learning environment and amortizing inventory
validation into the policy weights.

To achieve this without historical demand-side data, we formulate query generation as
a reinforcement learning problem where the live search engine acts as the environment
(Figure~\ref{fig:overview}). The LLM agent observes a state $c$ (the parameterized travel
context, such as destination and dates) and takes an action by generating a slate of
query suggestions $S$. These suggestions are executed against the environment (the
search engine), which retrieves physical inventory and returns a reward based on the
semantic quality of those listings, driving the policy update.

\textbf{Supply-Aware Reward Design.}
While our canonical evaluation metric, IMR@18, provides a high-fidelity assessment of
inventory, computing it during online training is computationally
intractable. It issues one judge call per query-listing pair, so with $G = 8$ sampled
slates per prompt, $|S| = 2$ suggestions per slate, and $K = 18$ retrieved listings per
suggestion, a single gradient step would block on $G \cdot |S| \cdot K = 288$
independent LLM calls. Across thousands of steps this is prohibitively slow and
expensive.

We therefore define $R_{18}$, a proxy read directly off the search engine's existing
reranker. Let $m(s_i, l_{i,j}) \in [0,1]$ be the reranker's query-listing semantic match
score for the $j$-th listing retrieved by suggestion $s_i$, a component that deliberately
isolates topical relevance from engagement and booking propensity. The slate-level reward is
the mean match over all retrieved listings:
{\scriptsize
\[
R_{18}(S) \;=\; \frac{1}{|S|\,K} \sum_{i=1}^{|S|} \sum_{j=1}^{K} m(s_i, l_{i,j}).
\]
}%
Crucially, this signal is free. The reranker already scores every listing it returns in the
serving path, so all $288$ per-listing scores arrive as a byproduct of the $G \cdot |S| = 16$
retrieval calls the rollouts make anyway, so the reward adds no model inference.

Over 9{,}000 suggestions across 300 contexts, $R_{18}$ and IMR@18 exhibit a stable
positive rank correlation (Spearman $\text{corr} = 0.45$, 95\% CI $[0.43, 0.48]$, bootstrapped
over contexts), justifying its use as a dense reward signal.

\textbf{Online Policy Update via GRPO.}
To actively explore the inventory distribution, SCOUT optimizes the policy $\pi_\theta$
using Group Relative Policy Optimization (GRPO)~\cite{shao2024grpo}. For each context
$c$, we sample a group of $G$ slates $\{S^{(1)}, \dots, S^{(G)}\}$. Each slate is
executed against the search engine to obtain rewards $\{r^{(1)}, \dots, r^{(G)}\}$, where
$r^{(i)} = R_{18}(S^{(i)})$. We compute a group-relative advantage $A^{(i)}$ by standardizing the
rewards within the group:
\begin{equation}
A^{(i)} = \frac{r^{(i)} - \mu}{\sigma}, \quad \text{where }
\mu = \frac{1}{G}\sum_{i=1}^G r^{(i)}, \enskip
\sigma = \sqrt{\frac{1}{G}\sum_{i=1}^G (r^{(i)} - \mu)^2}
\end{equation}
Using these advantages, SCOUT updates the policy by maximizing the following objective:
\begin{equation}
\label{eq:grpo}
\resizebox{\linewidth}{!}{$
\mathcal{J}_{GRPO}(\theta) = \mathbb{E}_{c, \{S^{(i)}\}_{i=1}^G} \left[ \frac{1}{G}
\sum_{i=1}^G \min \left( \rho^{(i)} A^{(i)}, \text{clip}(\rho^{(i)}, 1-\epsilon, 1+\epsilon) A^{(i)}
\right) - \beta \mathbb{D}_{KL}(\pi_\theta || \pi_{ref}) \right]
$}
\end{equation}
where $\rho^{(i)} = \frac{\pi_\theta(S^{(i)}|c)}{\pi_{\theta_{old}}(S^{(i)}|c)}$ is the probability
ratio, $\epsilon$ controls clipping, $\beta$ weights the KL penalty, and
$\pi_{ref}$ is the reference (base) policy.

\section{Results}
\label{sec:results}

\subsection{Experimental Setup}
\label{sec:results:setup}

\textbf{Setup.} All models are initialized from the Qwen3-8B base
model~\cite{qwen3} and trained using LoRA~\cite{hu2022lora} (rank = 32,
$\alpha = 64$, targeting all linear layers). To evaluate
generalization to unseen markets, the training and evaluation datasets are disjointly
split by destination. We evaluate all methods under the same setting used in training:
$|S| = 2$ suggestions per slate, $K=18$ listings per suggestion. Every trained
arm is held to the same training budget and evaluated across three random seeds; we
report the seed mean.

\noindent\textbf{Baselines.} We compare SCOUT against three paradigms:
\begin{itemize}[leftmargin=*, topsep=2pt, itemsep=2pt]
\item \textbf{Few-shot (Prompting):} The base model is explicitly instructed to
optimize for both diversity and inventory, provided with positive and
negative examples. It isolates what training adds over instruction.
\item \textbf{RSFT (Rejection-Sampling Fine-Tuning)~\cite{dong2023raft}:} We sample
$N=16$ slates per context from the Few-shot policy, score them using the $R_{18}$
metric, and retain the top-1 slate per context for supervised fine-tuning (SFT). RSFT
utilizes the identical reward signal as SCOUT but consumes it via imitation rather than
RL.
\item \textbf{DPO / RPO (Preference Optimization):} We construct preference pairs
$(y^+, y^-)$ such that the chosen response strictly improves inventory by a margin without
degrading diversity ($\Delta_{inv} \geq \delta_{inv}$ and $\Delta_{div} \geq 0$). We apply a
per-context cap to prevent any single context from dominating the pair set.
DPO~\cite{rafailov2023dpo} is trained with $\beta = 0.1$. RPO~\cite{liu2024rpo} utilizes
the same setup but introduces an auxiliary SFT loss ($\text{pref\_ftx} = 1.0$).
\end{itemize}

\textbf{SCOUT (Ours):} We run GRPO as described in \S\ref{sec:method:rl} with group size
$G=8$, KL penalty $\beta=0.06$, fully on-policy ($\texttt{num\_iterations}=1$),
initialized from the base model.

\begin{table}[t]
\centering
\small
\caption{Main results, relative to Few-shot for inventory and absolute for diversity
(three-seed means). SCOUT attains the best inventory grounding at parity diversity.
\textsuperscript{*}DPO's diversity collapse makes its inventory deltas not meaningful.}
\label{tab:main}
\begin{tabular}{l|cc|c}
\toprule
 & \multicolumn{2}{c|}{Inventory} & Diversity \\
 \cmidrule(lr){2-3}
Method & $\Delta R_{18}$ & $\Delta$IMR@18 & distinct\%@0.9 \\
\midrule
Few-shot & -- & -- & 84.03 \\
RSFT     & +5.24\% & +1.75\% & 84.11 \\
DPO      & n/a\textsuperscript{*} & n/a\textsuperscript{*} & 11.07 \\
RPO      & +18.12\% & +6.02\% & 84.04 \\
SCOUT    & +37.63\% & +12.30\% & 83.72 \\
\bottomrule
\end{tabular}
\end{table}

\subsection{Main Results}
\label{sec:results:main}

The performance of all methods is summarized in Table~\ref{tab:main}.

\textbf{The Limits of Prompting and SFT:} The Few-shot baseline achieves strong
diversity (84.03 distinct\%@0.9) but yields only modest inventory grounding.
This asymmetry defines the problem for every trained arm: raise
inventory-awareness, hold diversity. Prompting struggles because the model inherently
lacks access to the dynamic inventory of a given market. RSFT attempts to close this
gap by learning from best-of-$N$ generations, but yields only marginal gains
(+1.75\% IMR@18) because RSFT only copies winning samples: it never observes a
good-versus-bad contrast, and remains constrained by the frozen Few-shot policy's
output distribution.

\textbf{The Diversity Collapse in Preference Learning:} DPO explicitly learns from
contrastive pairs, but we observe a severe diversity collapse (11.07
distinct\%@0.9). The model concentrates its distribution on a narrow set of generic
queries due to likelihood displacement~\cite{razin2025displacement,ren2025squeezing}.
RPO mitigates this by anchoring the model with an auxiliary SFT loss. It restores
diversity to 84.04 and yields +18.12\% $R_{18}$ / +6.02\% IMR@18, showing the contrast is
informative once the collapse is mitigated.

\textbf{SCOUT Achieves a Balance:} SCOUT (GRPO) outperforms all baselines, achieving
the highest inventory-awareness (+37.63\% $R_{18}$, +12.30\% IMR@18) while maintaining
diversity. By optimizing over groups of generations rather than static
pairs, SCOUT explores the reward landscape and learns robust supply-aware
behaviors.

\subsection{Amortizing Serve-Time Search into Model Weights}
\label{sec:results:amortize}

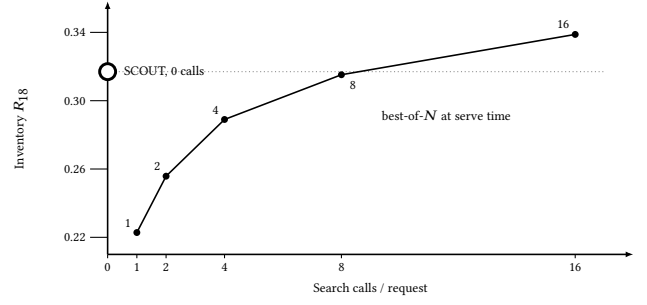
\begin{figure}[t]
\centering
\begin{tikzpicture}[x=11pt, y=90pt]
\draw[semithick,-{Latex[length=3pt]}] (0,0) -- (18,0);
\draw[semithick,-{Latex[length=3pt]}] (0,0) -- (0,1.06);
\foreach \yv/\lab in {0.07143/0.22, 0.35714/0.26, 0.64286/0.30, 0.92857/0.34}
  {\draw (0,\yv) -- (-0.6,\yv) node[left, font=\tiny, inner sep=1.5pt] {\lab};}
\foreach \xv/\lab in {0/0, 1/1, 2/2, 4/4, 8/8, 16/16}
  {\draw (\xv,0) -- (\xv,-0.016) node[below, font=\tiny, inner sep=1.5pt] {\lab};}
\node[below, font=\tiny] at (9,-0.085) {Search calls / request};
\node[rotate=90, font=\tiny, anchor=south] at (-2.45,0.53) {Inventory $R_{18}$};

\draw[densely dotted, black!45] (0,0.76429) -- (17,0.76429);
\draw[semithick] (1,0.09143) -- (2,0.32714) -- (4,0.56429) -- (8,0.75143)
  -- (16,0.92000);
\foreach \xv/\yv/\lab in {1/0.09143/1, 2/0.32714/2, 4/0.56429/4, 16/0.92000/16}
  {\fill (\xv,\yv) circle (1.3pt);
   \node[above left=-1pt and -1pt, font=\tiny] at (\xv,\yv) {\lab};}
\fill (8,0.75143) circle (1.3pt);
\node[below right=-1pt and 0pt, font=\tiny] at (8,0.75143) {8};
\node[right=1pt, font=\tiny] at (9,0.58) {best-of-$N$ at serve time};

\draw[line width=1.1pt, fill=white] (0,0.76429) circle (3pt);
\node[right=3pt, font=\tiny] at (0,0.76429) {SCOUT, 0 calls};
\end{tikzpicture}
\Description{Line chart of inventory reward R18 on the vertical axis against search calls per request on the horizontal axis. A best-of-N at serve time curve rises steadily from 1 to 16 calls. SCOUT is a single hollow point at 0 calls, with a dotted horizontal line at its level that sits just above the best-of-8 point.}
\caption{Inventory-awareness vs. serve-time cost. SCOUT achieves the inventory
grounding of a best-of-8 policy at zero marginal serve-time cost.}
\label{fig:amortize}
\end{figure}

Figure~\ref{fig:amortize} places these gains against the alternative of validating
inventory at serve time. Best-of-$N$ buys inventory-awareness with search calls; each
call is a full search request issued on every impression, so the cost recurs for the
lifetime of the feature. SCOUT pays it once during training instead: it reaches
$0.32$ with no search calls of its own, marginally above best-of-8 and $81\%$ of the
way from best-of-1 to best-of-16. This is what makes supply-aware query generation
deployable on a real-time travel search path.

\section{Conclusion and Ongoing Work}
\label{sec:conclusion}

We introduced SCOUT, a framework for generating supply-aware query
suggestions to bootstrap natural language interfaces. As a short paper on
ongoing work, our limitations and future roadmap are best understood through
a two-phase deployment strategy:

\textbf{Phase 1: System Capability Bootstrapping (Current).} SCOUT explicitly tackles
the cold-start problem by aligning LLMs with supply-side constraints. Consequently,
our current limitations include a reliance on offline evaluation metrics and a
reward signal that scores fulfillability but not appeal. However, this pure supply-aware
bootstrapping is a necessary precondition; it ensures the initial system is reliable
and functional enough to launch, which is required to collect meaningful user
interaction data.

\textbf{Phase 2: Demand-Aware Alignment (Ongoing).} An ideal query suggestion system
must synthesize both supply availability and demand preferences. As the interface
rolls out, our ongoing work focuses on large-scale online A/B testing to capture live
demand-side signals (e.g., click-through rates, booking conversions). This data will
drive our transition from pure supply-awareness to a balanced, multi-objective
alignment strategy, ensuring future suggestions are physically fulfillable and
resonant with user intent.

\label{endbody}   

\balance

\bibliographystyle{ACM-Reference-Format}
\bibliography{refs}

\end{document}